\documentclass[aps,pra,twocolumn,superscriptaddress]{revtex4-2}
\usepackage[T1]{fontenc}
\usepackage[utf8]{inputenc}
\usepackage{color}
\usepackage{babel}
\usepackage{amsmath}
\usepackage{graphicx}
\usepackage{comment}
\usepackage{esint}
\usepackage[unicode=true,pdfusetitle,
 bookmarks=true,bookmarksnumbered=false,bookmarksopen=false,
 breaklinks=false,pdfborder={0 0 0},pdfborderstyle={},backref=false,colorlinks=true]
 {hyperref}

\makeatletter

\usepackage{hyperref}
\hypersetup{
   pdfpagemode=None,
   pdfstartpage=1,
   pdfmenubar=true,
   pdftoolbar=true,
   colorlinks = true,
   linkcolor=blue,
   citecolor=blue,
   urlcolor=blue,
   bookmarksopen=false
 }

\usepackage{dcolumn}
\usepackage{bm}
\usepackage{color}
\usepackage{soul}
\usepackage{babel}
\usepackage{amsfonts}
\usepackage{slashed}
\usepackage{enumerate}
\usepackage{subfigure}
\usepackage{mathtools}
\usepackage{float}

\usepackage{babel}

\makeatother

\newcommand{\bra}[1]{\left\langle #1 \right|}
\newcommand{\ket}[1]{\left| #1 \right\rangle}
\newcommand{\braket}[2]{\bra{#1} #2\rangle}

\usepackage{xspace}
\newcommand{\Cite}{\unskip~\cite}

\begin{document}

\title{Nonlocal transfer of high-dimensional unitary operations}
\author{Dilip Paneru}
\affiliation{Nexus for Quantum Technologies, University of Ottawa, K1N 5N6, Ottawa, ON, Canada}
\author{Francesco Di Colandrea}
\email[Correspondence email address: ]{francesco.dicolandrea@unina.it}
\affiliation{Nexus for Quantum Technologies, University of Ottawa, K1N 5N6, Ottawa, ON, Canada}
\affiliation{Dipartimento di Fisica, Universit\`{a} degli Studi di Napoli Federico II, Complesso Universitario di Monte Sant'Angelo, Via Cintia, 80126 Napoli, Italy}
\author{Alessio D'Errico}
\affiliation{Nexus for Quantum Technologies, University of Ottawa, K1N 5N6, Ottawa, ON, Canada}
\affiliation{National Research Council of Canada, 100 Sussex Drive, Ottawa ON Canada, K1A 0R6}
\author{Ebrahim Karimi}
\affiliation{Nexus for Quantum Technologies, University of Ottawa, K1N 5N6, Ottawa, ON, Canada}
\affiliation{National Research Council of Canada, 100 Sussex Drive, Ottawa ON Canada, K1A 0R6}
\affiliation{Institute for Quantum Studies, Chapman University, Orange, California 92866, USA}

\begin{abstract}
Highly correlated biphoton states are powerful resources in quantum optics, both for fundamental tests of the theory and practical applications. In particular, high-dimensional spatial correlation has been used in several quantum information processing and sensing tasks, for instance, in ghost imaging experiments along with several quantum key distribution protocols. Here, we introduce a technique that exploits spatial correlations, whereby one can nonlocally access the result of an arbitrary unitary operator on an arbitrary input state without the need to perform any operation themselves. The method is experimentally validated on a set of spatially periodic unitary operations in one-dimensional and two-dimensional spaces. Our findings pave the way for efficiently distributing quantum simulations and computations in future instances of quantum networks where users with limited resources can nonlocally access the results of complex unitary transformations via a centrally located quantum processor. 
\end{abstract}
\maketitle

\section*{Introduction}
Quantum entanglement~\cite{horodecki2009quantum,paneru2020entanglement}, one of the central concepts of quantum mechanics, has led to fundamental modifications to our understanding of the physical world~\cite{einstein1935can,bell1964einstein}, along with exciting technological developments in computation~\cite{jozsa2003role}, metrology~\cite{giovannetti2011advances,demkowicz2014using}, and communication~\cite{bennett2014quantum,ursin2007entanglement}. One of the most widely used techniques to generate entangled photon pairs exploits a nonlinear process known as spontaneous parametric down-conversion (SPDC)~\cite{hong1985theory}, whereby the emitted photons can be highly correlated in polarization, frequency, and spatial degrees of freedom~\cite{walborn2010spatial}. Down-converted photon pairs have been employed as a source of polarization entanglement for fundamental tests of nonlocality~\cite{aspect1982experimental}, in quantum teleportation~\cite{bouwmeester1997experimental}, and communication protocols~\cite{ursin2007entanglement}.
The high-dimensional correlation in the position degree of freedom (and the corresponding anti-correlation in the momentum space) can be exploited in several quantum imaging experiments~\cite{moreau2019imaging}, such as quantum ghost imaging~\cite{shapiro2012physics} and biphoton holography~\cite{devaux2019quantum,zia2023interferometric}.

When two parties share entangled particles, either can exploit the quantum correlations to nonlocally affect the quantum state of the other—for instance, through local measurements—a technique also referred to as quantum steering~\cite{Schrödinger_1935,Schrödinger_1936,PhysRevLett.98.140402}. This concept was originally proposed and demonstrated for two photons entangled in the polarization degree of freedom, i.e., two photonic qubits~\cite{Huelga_2005}. However, the extension to high-dimensional entangled states unlocks remarkable advantages for quantum information protocols, in particular a higher information capacity and increased noise tolerance \Cite{PhysRevA.61.062308}. The first experimental demonstration of high-dimensional steering was reported in Ref.~\cite{PhysRevLett.126.200404} with photons entangled in the discretized position-momentum degree of freedom \Cite{HerreraValencia2020highdimensional}, up to 31 dimensions.

Entanglement also enables the implementation of teleportation-based quantum computation, also known as gate teleportation \Cite{Gottesman1999}, consisting of applying a quantum gate to an entangled state and then teleporting the target state through it \Cite{Bouwmeester1997}.~This concept has been thoroughly investigated theoretically~\cite{Caha2024singlequbitgate,PhysRevApplied.23.014064} and demonstrated experimentally for qubit systems \Cite{PhysRevLett.93.240501,doi:10.1073/pnas.1005720107,Chou2018,doi:10.1126/science.aaw9415}.~Recently, schemes to extend the technique to higher-dimensional (qudit) systems have also been proposed \Cite{high-dim-gate_teleport}.

In this paper, we show how the inherent high degree of spatial correlations between photon pairs can be exploited to also obtain a protocol for transferring the output of a unitary operation performed on one of the photons (signal) to the second (idler). Experimentally, the optical transformation is implemented via a Spatial Light Modulator (SLM) encoding several phase masks. In combination with a multimode fiber, a similar setup has recently been adopted as a programmable photonic circuit processing spatial modes~\cite{goel2024inverse}, with the outcome of the transformed photon revealed via projective measurements performed on the correlated one. In our experiment, the masks are generated as superpositions of sinusoidal gratings that mimic the result of lattice dynamics on optical modes that carry a quantized amount of transverse momentum, recently introduced for the simulation of discrete-time quantum walk dynamics~\cite{DErrico2020,DiColandrea2023}. Our method is validated on a set of unitary operations coupling single input modes into multiple output modes, in both one-dimensional (1D) and two-dimensional (2D) configurations. In the 1D realization, a practical method for engineering the simulation transfer for different state preparations is also illustrated. 

The versatility and robustness of the protocol suggest its employment in a photonic quantum network where computational capabilities are centralized. In such a framework, we envision a party with access to a quantum simulator that can perform the required operations, while remote clients without direct access to the platform can securely retrieve the simulation output. This approach paves the way for alternative implementations of blind quantum computation~\cite{Fitzsimons2017}, as well as for distributing quantum simulations across different nodes of quantum networks~\cite{review_network}, ultimately enabling resource-efficient and scalable quantum computing solutions.

\section*{Theory}
\label{sec:theory}
The biphoton wavefunction generated from Type-I degenerate SPDC from thin crystals can be expressed as
\begin{equation}
   \ket{\psi} = \mathcal{C} \int \text{d}\textbf{k}\, \ket{\textbf{k}}_s\ket{-\textbf{k}}_i,
   \label{eqn:spdc_far}
\end{equation}
where $\textbf{k}$ is the transverse momentum, $C$ a normalization factor, and $s$ and $i$ refer to signal and idler photon, respectively. Equation~\eqref{eqn:spdc_far} expresses the momentum conservation for the SPDC process and assumes a plane-wave pump. 


We aim to define a scheme to transfer the result of a unitary transformation $\hat{U}$, acting only on the signal photon, to the idler photon. In the following, we show that this can be achieved by successive application of a different unitary operation and a projective measurement on the signal. The action of $\hat{U}$ on a momentum state $\ket{\textbf{k}}$ is defined as
\begin{eqnarray}
    \hat{U}\ket{\textbf{k}} = \int \text{d}\textbf{k}'\,  U(\textbf{k}',\textbf{k}) \ket{\textbf{k}'}.
    \label{nstepunitary}
\end{eqnarray}

Our protocol for successfully transferring this operation from the \emph{signal} to the \emph{idler} photons relies on applying a different unitary $\hat{U}'_s$ to the signal photon, followed by a suitable projection. This allows us to access the result of the unitary operator on any input state on the idler side, without performing any operation on the idler photons. The construction of $\hat{U}'_s$ and the required projective measurement are explained in further detail below. 
%
Assume we want to transfer the result of $\hat{U}$ on a single input mode $\ket{\phi_0}=\ket{\textbf{k}_0}$. Upon applying the unitary $\hat{U}'_s$ to the signal photon, the biphoton state is transformed to
\begin{eqnarray}
   \ket{\psi'} &=& \hat{U}'_s \otimes \hat{I} \ket{\psi} 
   \label{eqn:unitaryonspdc}  
   \\
               &=& \iint \text{d}\textbf{k}'\text{d}\textbf{k}\,  U'_s(\textbf{k}',\textbf{k})  \ket{-\textbf{k}}_i\ket{\textbf{k}'}_s, 
               \nonumber
\end{eqnarray}
where $\hat{I}$ is the identity operator in the idler basis.
If a projection $\Pi_{\textbf{k}'_0}$ on $\ket{\textbf{k}'_0}_s$ is performed, the state of the idler photon is transformed according to
\begin{eqnarray}
    \ket{\phi}_i &=&  \int \text{d}\textbf{k}' \left(\int \text{d}\textbf{k} \, U'_s(\textbf{k}',\textbf{k})  \ket{-\textbf{k}}_i\right) \braket{\textbf{k}'_0}{\textbf{k}'}_s \label{eqn:stateafterproj}\\
    &=& \int \text{d}\textbf{k} \, U'_s(\textbf{k}'_0,-\textbf{k})  \ket{\textbf{k}}_i,
    \nonumber
\end{eqnarray}
where we used $\braket{\textbf{k}'_0}{\textbf{k}'}=\delta(\textbf{k}'_0-\textbf{k}')$, with $\delta(\textbf{x})$ the Dirac delta function.
By comparing the final result with Eq.~\eqref{nstepunitary}, one obtains that the target unitary operator $\hat{U}_s'$ can be constructed as
\begin{equation}
    U'_s(\textbf{k}',\textbf{k}) = U(-\textbf{k},\textbf{k}'),
    \label{eqn:unitaryconstruction}
\end{equation}
and the result for a localized input state $\ket{\textbf{k}_0}$ can be accessed simply by projecting on $\ket{\textbf{k}_0'}=\ket{\textbf{k}_0}$.

For the more general case of transferring the result of a unitary operation on an arbitrary initial state, which generally includes a superposition of modes, one can still apply a carefully chosen unitary operator $\hat{U}'$ to the signal photon, followed by a projection on a suitable state $\ket{\chi}_s$. The generalization is demonstrated in the following. 
A general input state can be written as
\begin{eqnarray}
    \ket{\phi_0}_i = \int \text{d}\textbf{k}'\, C(\textbf{k}') \ket{\textbf{k}'}_i,
\end{eqnarray}
where $C(\textbf{k}') $ is the coefficient of the momentum mode $\ket{\textbf{k}'}_i$. 
The result of the unitary operator on such an initial state can be written as
\begin{eqnarray}
  \hat{U}\ket{\phi_0} &=& \int d \textbf{k} \left( \int d\textbf{k}'C(\textbf{k}') U(\textbf{k},\textbf{k}')\right) \ket{\textbf{k}}.
  \label{unitaryongeneralinitialstate}
\end{eqnarray}
Let us also define the general state for projection $\ket{\chi}_s$ as
\begin{eqnarray}
 \ket{\chi}_s = \int d\textbf{k}'\, A(\textbf{k}') \ket{\textbf{k}'}_s. 
\end{eqnarray}
As in Eqs.~\eqref{eqn:unitaryonspdc} and \eqref{eqn:stateafterproj}, upon the application of the unitary $\hat{U}'_s$ to the signal photon and a projection on $\ket{\chi}$, the idler state transforms as
\begin{eqnarray}
\ket{\phi}_i &=& \int \text{d}\textbf{k}' \left( \int 
                 \text{d}\textbf{k}\,  U'_s(\textbf{k}',\textbf{k})  \ket{-\textbf{k}}_i \right) \braket{\chi}{\textbf{k}'}_s  \nonumber \cr
            &=&   \int \text{d}\textbf{k} \left(\int 
                 \text{d}\textbf{k}'\, A^*(\textbf{k}') U'_s(\textbf{k}',-\textbf{k})   \right) \ket{\textbf{k}}_i.
\end{eqnarray}
By comparing the final result with Eq.~\eqref{unitaryongeneralinitialstate}, and fixing $\hat{U}'_s$ as in Eq.~\eqref{eqn:unitaryconstruction}, we obtain the coefficients $A(\textbf{k}')$ for the required projection: 
%
%
\begin{eqnarray}
     A(\textbf{k}') &=& \left(\frac{C(\textbf{k}') U(\textbf{k},\textbf{k}')}{U'_s(\textbf{k}',-\textbf{k})}\right)^*  \label{eqn:generalcoeffs}\\
     &=& C(\textbf{k}')^*.
     \nonumber
\end{eqnarray}

\begin{figure}[t]
  \centering
  \includegraphics[scale=0.15]{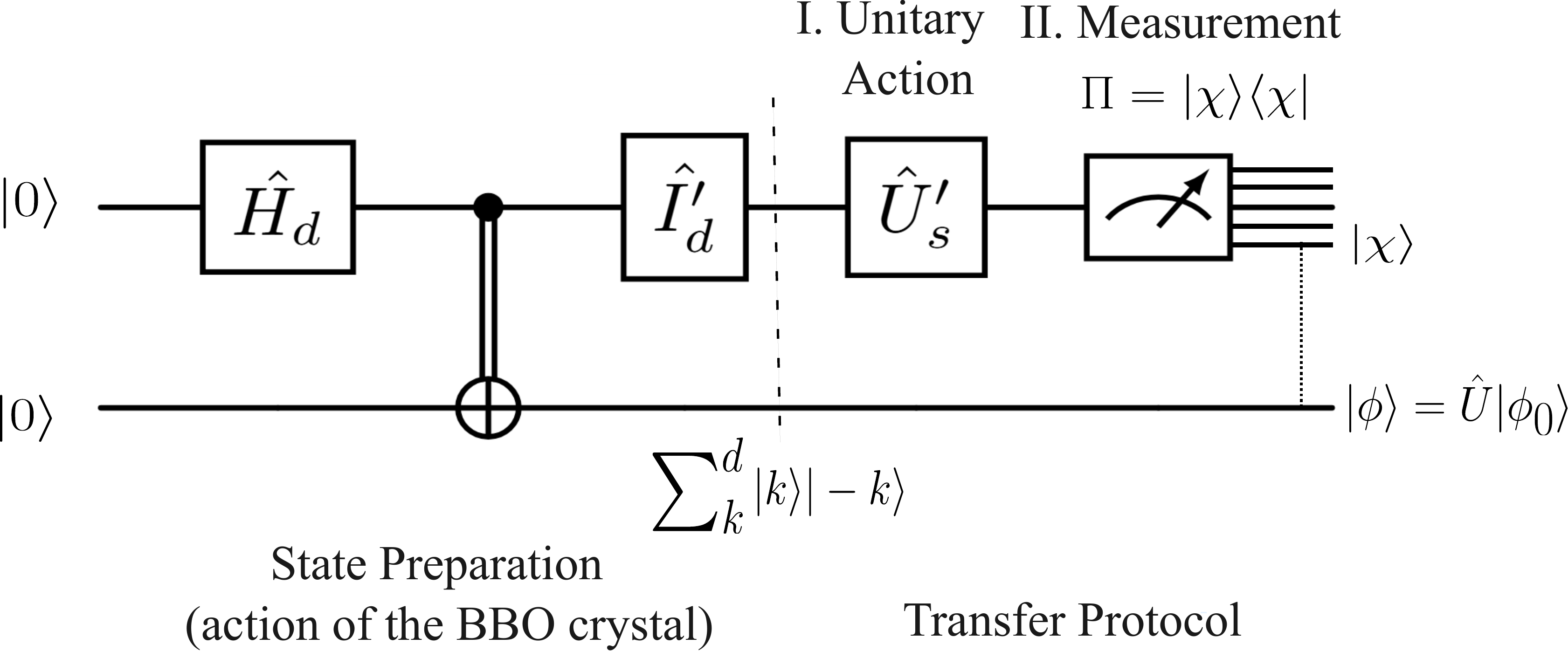}
  \caption{\textbf{Quantum circuit of the nonlocal transfer.}
  The first part of the circuit represents the state preparation, which involves the application of a $d$-dimensional Hadamard operator, followed by a $d$-dimensional control X gate with an action $\ket{j}\ket{k} \rightarrow \ket{k}\ket{(j+k)\, \text{Mod  $d$}}$ followed by the $d$-dimensional Anti-Identity operator which creates the state $\sum_k\ket{k}\ket{-k}$. In our experiment, the action of the pump on the BBO crystal already prepares the desired state. The second part consists of the unitary action ${\hat{U}_s}$ on the signal photon, followed by a projective measurement $\Pi$. Upon successful projection, the desired unitary action $\hat{U}$ on the desired state $\ket{\phi_0}$ is obtained on the idler photon.}
  \label{Fig:CircuitDiagram}
\end{figure}

Thus, if the idler party needs to simulate the action of $\hat{U}$ on an arbitrary input state ${\ket{\phi_0}}$ but has no access to any computational resource, the signal party can perform $\hat{U}_s'$, as prescribed by Eq.~\eqref{eqn:unitaryconstruction}, followed by a projection on the state $\ket{\chi}_s$, whose coefficients are given by Eq.~\eqref{eqn:generalcoeffs}. In this way, the resource for the unitary operation remains centralized, but the results can be distributed across multiple parties in a network, effectively trading off high-dimensional correlations. We show the conceptual picture of the protocol in the form of a quantum circuit in Fig.~\ref{Fig:CircuitDiagram}.
%
\begin{figure*}[t!]
  \centering
  \includegraphics[scale=0.5]{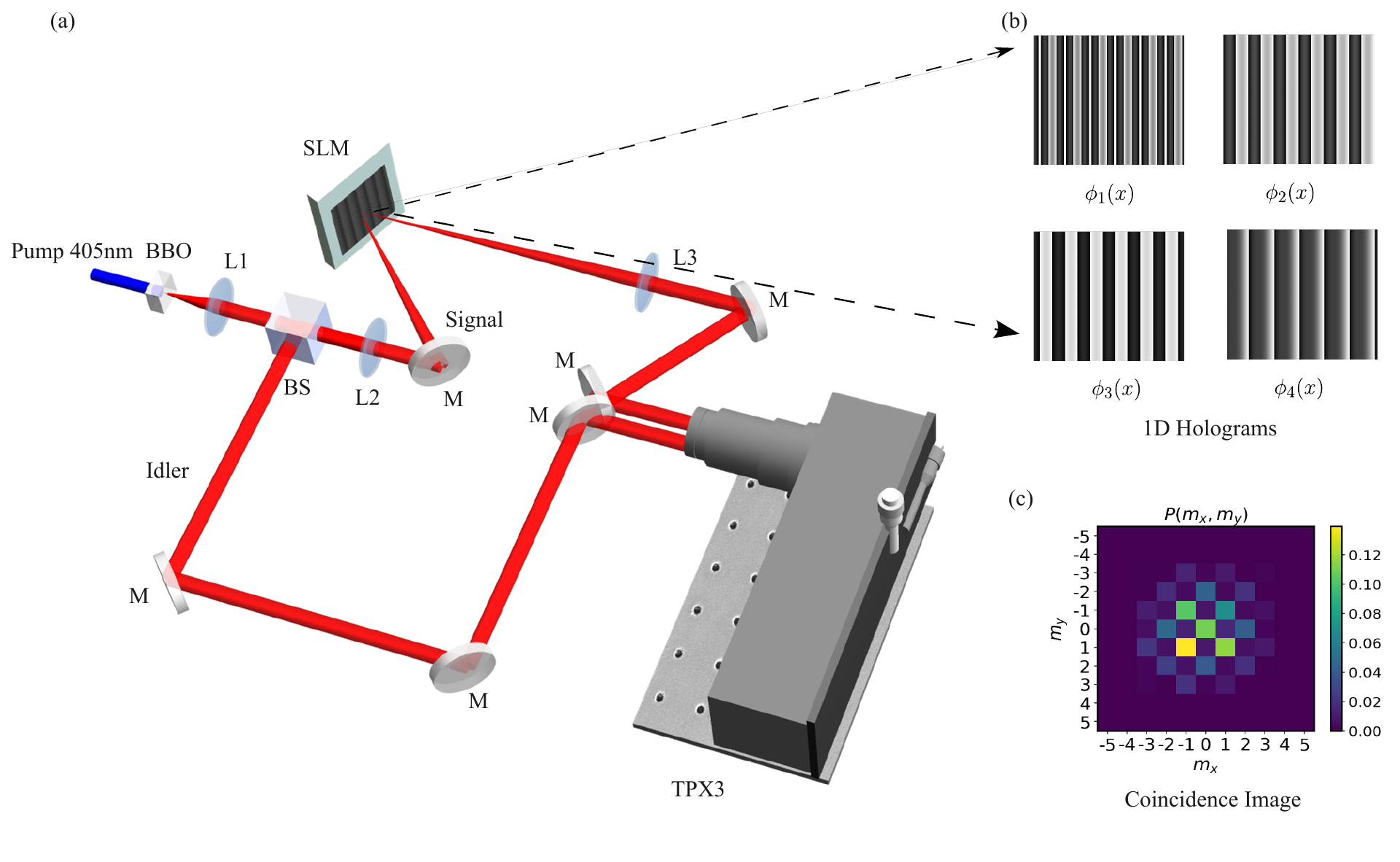}
  \caption{\textbf{Experimental setup.} (a)~A 405 nm laser illuminates a 1-mm type-I BBO crystal generating degenerate down-converted photon pairs. The idler and signal photons are probabilistically separated by a 50:50 beamsplitter (BS). The phase profile of the signal photon is modified by a Spatial Light Modulator (SLM), placed in the image plane of the crystal. The signal and idler photons are imaged onto different regions of a time-tagging TPX3CAM camera (TPX3), placed in the far field of the holograms. By performing a suitable postselection of the events in the signal image, the result of the unitary operation can be transferred to the idler photons. 
  (b)~Holograms for phases, $\phi_1(x)=1.3\sin(\Delta k_{\perp}x)+1.5\cos(2\Delta k_{\perp}x)$, $\phi_2(x)=1.9\sin(\Delta k_{\perp}x)$,  $\phi_3(x)=\cos(\Delta k_{\perp}x)$, and ${\phi_4(x)=\cos(\Delta k_{\perp}x)+\Delta k_{\perp}x}$, shown in grayscale color, corresponding to 1D unitary operations. The transferred far-field distribution recorded on the idler photon is shown in panel (c) for a 2D unitary as an example. BBO: Beta Barium Borate crystal; BS: Beamsplitter; SLM: Spatial Light Modulator; L1, L2, and L3: Lenses; M:Mirror. }
  \label{fig:Setup}
\end{figure*}

For the purpose of experimental demonstration, we implement unitary transformations in the form of phase masks with an SLM. The chosen unitaries correspond to the simulation of lattice dynamics on optical modes carrying a quantized amount of transverse momentum, generated from the superpositions of sinusoidal gratings.
Such operators act on the transverse momentum degree of freedom of the signal photon, adding quantized amounts of transverse momentum, ${\Delta k_{\perp} = 2\pi/\Lambda }$, where $\Lambda$ is a characteristic distance. The action of $\hat{U}$ on a momentum state can therefore be expressed as
\begin{equation}
\hat{U}\ket{k}_s=\sum_{m} u_{m,k}\ket{m_k}_s,
\label{eqn:unitary}
\end{equation}
where ${\ket{m_k}=\ket{k+k_m}}$, with ${k_m=m\Delta k_\perp}$ and $m$ an integer number. If translation invariance is assumed, then ${u_{m,k}=u_m}$. The resulting biphoton state can be written as
\begin{equation}
\ket{\psi'}=\sum_m\int \text{d}k\,u_{m}\ket{m_k}_s\ket{-k}_i.
\end{equation}

If a projection is performed onto a specific signal state, say $\ket{k_0}_s$, the obtained state on the idler side is
\begin{equation}
\ket{\psi}_i = \sum_m u_{m}\ket{k_m-k_0}_i.
\label{eqn:fundamentallattice}
\end{equation}

\begin{figure*}[t!]
  \centering
  \includegraphics[scale=0.46]{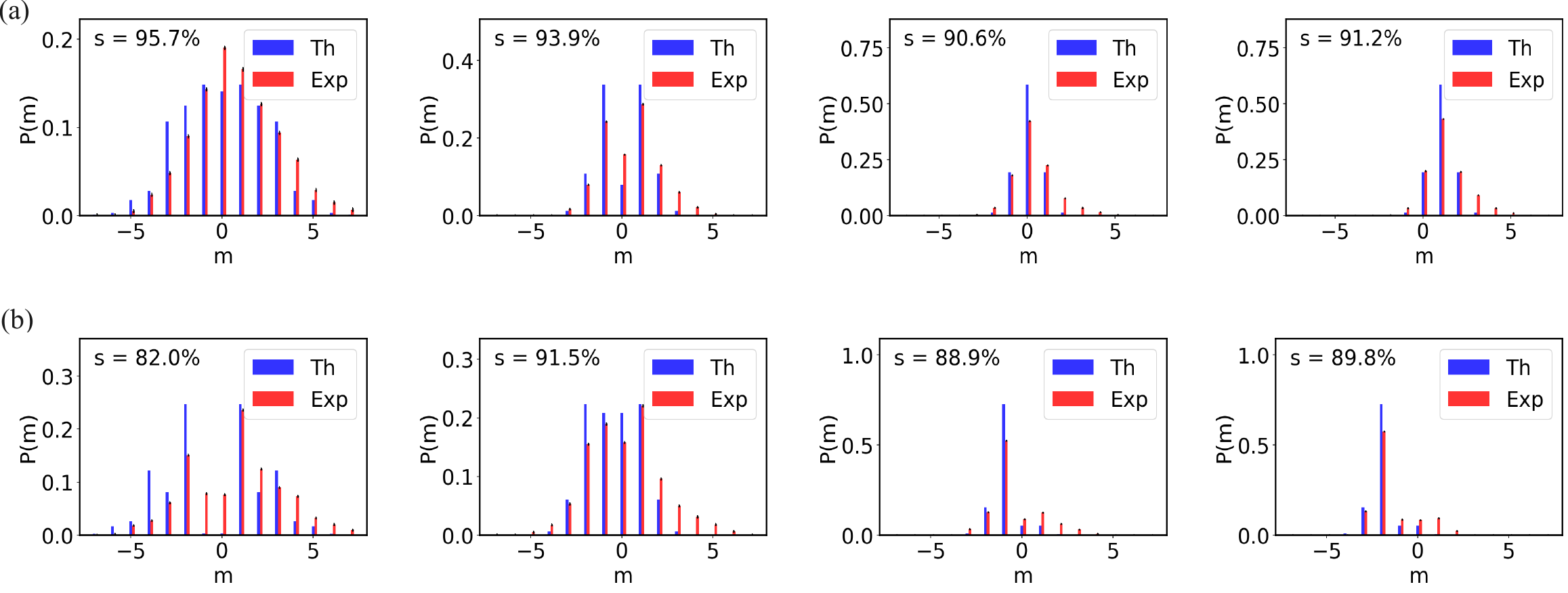}
  \caption{\textbf{One-dimensional nonlocal transfer.} (a)~Experimental probabilities for the momentum modes of the idler photon upon nonlocal transfer of different 1D lattice unitary operations, compared with theoretical predictions. From left to right: $\phi_1(x)=1.3\sin(\Delta k_{\perp}x)+1.5\cos(2\Delta k_{\perp}x)$, $\phi_2(x)=1.9\sin(\Delta k_{\perp}x)$, $\phi_3(x)=\cos(\Delta k_{\perp}x)$, and ${\phi_4(x)=\cos(\Delta k_{\perp}x)+\Delta k_{\perp}x}$.  (b)~One-dimensional nonlocal transfer with arbitrary input states. Theoretical and experimental probabilities for the momentum modes of the idler photon upon nonlocal transfer of the same unitaries acting on the initial state ${\ket{\psi_0}=(\ket{0}+i\ket{1})/\sqrt{2}}$.}
  \label{Fig:1Dwalk}
\end{figure*}
\section*{RESULTS}
The experimental setup is sketched in Fig.~\ref{fig:Setup}. It consists of a 1-mm thick Type-I BBO ($\beta$-Barium Borate) crystal pumped by a 405~nm pulsed laser. Down-converted signal and idler photons 
are generated and then probabilistically separated into two paths through a beamsplitter (BS). Two lenses in a ${4f}$ configuration are used to image the crystal plane onto the SLM plane on the signal side. A half-wave plate (not shown in the figure) is used to rotate the signal input polarization in order to maximize the conversion efficiency from the SLM. A phase hologram $\phi(x)$ corresponding to a particular unitary process is displayed on the SLM. The action of such a hologram on the signal photon can be expressed in the position basis as
\begin{equation}
\ket{k}_s\rightarrow \int \text{d}x\, e^{i(\phi(x)+kx)}\ket{x}_s.
\label{eqn:nearfieldSLM2}
\end{equation}
Equation~\eqref{eqn:nearfieldSLM2} corresponds to the unitary action of the operator $\hat{U}$ visualized in the position space (cf.~Eq.~\eqref{eqn:unitary}). The holograms are generated as superpositions of sinusoidal gratings featuring spatial frequencies that are multiples of the transverse 
momentum unit ${2\pi/\Lambda}$. In our experiment, we set ${\Lambda=1\,\text{mm}/7\simeq 0.15\,\text{mm}}$. Afterward, both the signal and idler photons are redirected to two different regions of a time-stamping camera (TPX3CAM)~\cite{timepix1,timepix2}. The camera is placed in the far field of the nonlinear crystal, which allows us to access the transverse momentum space of the signal and the idler photons. This allows us to experimentally verify the plane-wave pump approximation assumed in Eq.~\eqref{eqn:spdc_far} by recording the momentum correlations of the photon pairs. These show very sharp, 1-pixel-wide anti-correlations, which validates the approximation.

The camera allows us to observe space-resolved coincidence images between the signal and the idler photons. In this way, upon postselection of a specific signal state, $\ket{k_0}_s$, the excited spectrum of momentum modes can be revealed in the far field of the idler photon,  
\begin{equation}
\ket{\psi}_i= \sum_{k_m} u_m\ket{k_m-k_0}_i, 
\label{eqn:farfieldSLM1}
\end{equation}
where each coefficient $u_m$ corresponds to the $m$-th element of the phase function in the Fourier basis:
\begin{equation}
u_m=\int \text{d}x\,e^{i\phi(x)}e^{-ik_m x}.   
\label{eqn:probdistr}
\end{equation}
With the imaging system used in our setup, each momentum mode covers an area of approximately five pixels at the imaging plane.
This scheme is used to implement a nonlocal transfer of the output of different unitary operations in both 1D and 2D configurations.

\subsection{One-dimensional simulation}
The experimental results obtained for different 1D unitary operators are shown in Fig.~\ref{Fig:1Dwalk}(a). The projection on the signal state $\ket{k_0=0}_s$ is chosen for reference. As discussed above, upon suitable postselection of the signal events, the idler far-field distribution is discretized, and a normalized spectrum of momentum modes $P(m)$ is extracted. The latter can be interpreted as the probability of occupation of the lattice sites spanned by the optical modes introduced in Eq.~\eqref{eqn:unitary}. A comparison with the theoretical predictions, extracted from Eq.~\eqref{eqn:probdistr}, is also provided. The agreement with the experimental observations is quantified by the similarity estimator $s =( \sum_m \sqrt{P_\text{exp}(m)P_\text{th}(m)}  )^2$, where ${P_\text{exp}(m)}$ and ${P_\text{th}(m)}$ are the experimental and theoretical far-field distributions, respectively. For all realizations, the similarity value exceeds 90\%, which showcases the accuracy of our method and its robustness to experimental imperfections, such as a residual misalignment of the input polarization state with respect to the SLM optic axis, which results in higher contributions from the zeroth diffraction order. Another issue limiting the performance of our apparatus is the low resolution of the SLM employed in the experiment, (${600\times 792}$) pixels. Poissonian statistics is assumed for computing error bars, which are always smaller than data points. Specifically, the holograms prepared in the 1D experiment are ${\phi_1(x)=1.3\sin(\Delta k_{\perp}x)+1.5\cos(2\Delta k_{\perp}x)}$, ${\phi_2(x)=1.9\sin(\Delta k_{\perp}x)}$, ${\phi_3(x)=\cos(\Delta k_{\perp}x)}$, and $\phi_4(x)=\cos(\Delta k_{\perp}x)+\Delta k_{\perp}x$. Note that the transformation induced by $\phi_4(x)$ is equivalent to $\phi_3(x)$ with an additional initial one-site displacement. The corresponding similarities are ${s= 95.7\%, 93.9\%, 90.6\%}$, and ${91.2\%}$. 

%
\begin{figure}[t]
  \centering
  \includegraphics[scale=0.48]{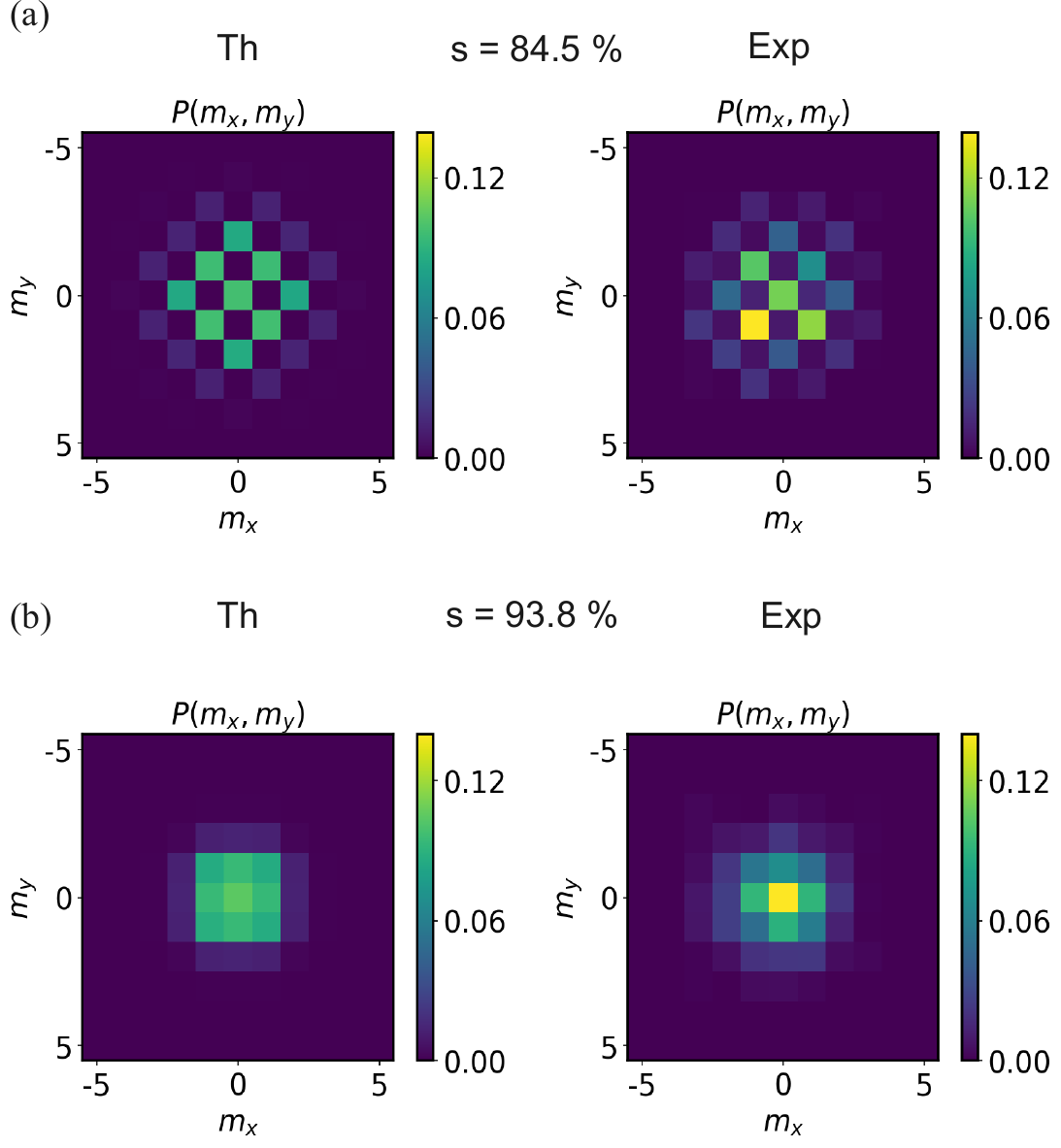}
  \caption{\textbf{Two-dimensional nonlocal transfer.} Theoretical and experimental probabilities for the momentum modes of the idler photon upon nonlocal transfer of different 2D lattice unitaries. (a)~${\phi_1(x,y)=2.8\sin{\left(\Delta k_\perp x\right)}\cos{\left(\Delta k_\perp y\right)}}$ and (b)~${\phi_2(x,y)=1.4\sin{\left(\Delta k_\perp x\right)}+1.4\sin{\left(\Delta k_\perp y\right)}}$.} 
  \label{Fig:2Dexperiment}
\end{figure}
%

The 1D implementation also allows us to implement the transfer of a unitary action on an arbitrary input state. Assume that the idler party requests computing the output of the unitary $\hat{U}$ on a general input state
\begin{equation}
\ket{\psi_0}=\sum_\ell d_\ell \ket{k_\ell},
\end{equation}
where $d_\ell$ are complex coefficients obeying the normalization condition ${\sum_\ell |d_\ell|^2=1}$. Accordingly, the target state
\begin{equation}
\ket{\psi}_i=\sum_n \sum_\ell u_nd_\ell\ket{k_\ell+k_n}_i 
\label{eqn:idleri}
\end{equation}
is expected to be transferred from the signal end. To accomplish this, the signal photon is sent through a different operator $\hat{\mathcal{V}}$:
\begin{equation}
\ket{\psi}=\sum_m\int\text{d}k\,v_m \ket{k+k_m}_s\ket{-k}_i,
\end{equation}
where $v_m$ are the elements of the spatial transformation associated with $\hat{\mathcal{V}}$ in the Fourier basis. Upon projection on a signal state, say $\ket{k_0=0}_s$, the resulting idler state reads
\begin{equation}
\ket{\psi}_i=\sum_m v_m\ket{k_m}_i.
\label{eqn:idlerf}
\end{equation}
Therefore, the specific operator $\hat{\mathcal{V}}$ to apply on the signal photon can be determined by equating Eq.~\eqref{eqn:idleri} and Eq.~\eqref{eqn:idlerf}, which yields
\begin{equation}
v_m=\sum_\ell d_\ell u_{m-\ell}.
\label{eqn:masking}
\end{equation}
The last equation reveals that the required operation is not a mere phase transformation. To effectively realize this transformation with a phase-only SLM, we employ the technique introduced in Ref.~\cite{Bolduc:13}, which enables the manipulation of both phase and amplitude of the field with a single phase-only hologram at the cost of introducing additional losses. In particular, the Fourier transform of the desired field is found in correspondence with the first diffraction order. For this reason, the required phase-amplitude transformation, extracted from Eq.~\eqref{eqn:masking}, is applied along $x$, and a blazing function ${\text{Mod}(2\pi/\Lambda_y,2\pi)}$ is added along the $y$ direction, with ${\Lambda_y=\Lambda/50=0.02\,\text{mm}}$. The spatial period $\Lambda_y$ is chosen to be small enough to ensure a clear separation between the first diffraction order and the unmodulated light in the Fourier plane. 

As a representative example, we apply this technique to transfer the outcomes of the same transformations considered before when applied to the delocalized input state ${\ket{\psi_0}=\left(\ket{0}+i\ket{1} \right)/\sqrt{2}}$. The experimental results are shown in Fig.~\ref{Fig:1Dwalk}(b). Good agreement with the theoretical distribution is observed, with an average similarity of $88.1\%$. This demonstrates the possibility of transferring also unitary operations that, in our optical encoding, do not correspond to simple phase transformations.

\subsection{Two-dimensional simulation}
The same concept is also tested in a 2D setting, where Eqs.~\eqref{eqn:farfieldSLM1} and ~\eqref{eqn:probdistr} are generalized as follows:
\begin{subequations}
\begin{equation}
\ket{\psi}_i= \sum_{k_{mx}}\sum_{k_{my}} u_{m_x,m_y}\ket{k_{mx}-k_{0x},k_{my}-k_{0y}}_i\,;
\end{equation}
\begin{equation}
u_{m_x,m_y}=\iint \text{d}x\,\text{d}y\,e^{i\phi(x,y)}e^{-ik_{mx} x}e^{-ik_{my} y}.
\end{equation}
\end{subequations}
The experimental results obtained for the 2D implementation are shown in Fig.~\ref{Fig:2Dexperiment}, obtained upon postselection of the signal state ${\ket{k_{0x},k_{0y}}_s=\ket{0,0}_s}$. In particular, two different simulations are considered, ${\phi_1(x,y)=2.8\sin{\left(\Delta k_\perp x\right)}\cos{\left(\Delta k_\perp y\right)}}$ and ${\phi_2(x,y)=1.4\sin{\left(\Delta k_\perp x\right)}+1.4\sin{\left(\Delta k_\perp y\right)}}$. The recorded similarities are ${s=84.5\%,}$ \text{and} ${93.8\%}$, respectively. 
Notably, in the 2D case, the number of active modes grows faster than in the 1D implementation, thus allowing us to access large-scale simulations with fewer computational resources. Specifically, while in the 1D case the number of accessible modes scales linearly with the number of spatial frequencies encoded in the phase masks, in the 2D case this scaling becomes quadratic. In other words, by increasing the number of spatial frequencies in the holograms, the 2D configuration can support a significantly larger Hilbert space, enhancing the system's capability to simulate more complex quantum dynamics or encode high-dimensional quantum information.



\begin{figure}[t]
  \centering
  \includegraphics[scale=0.27]{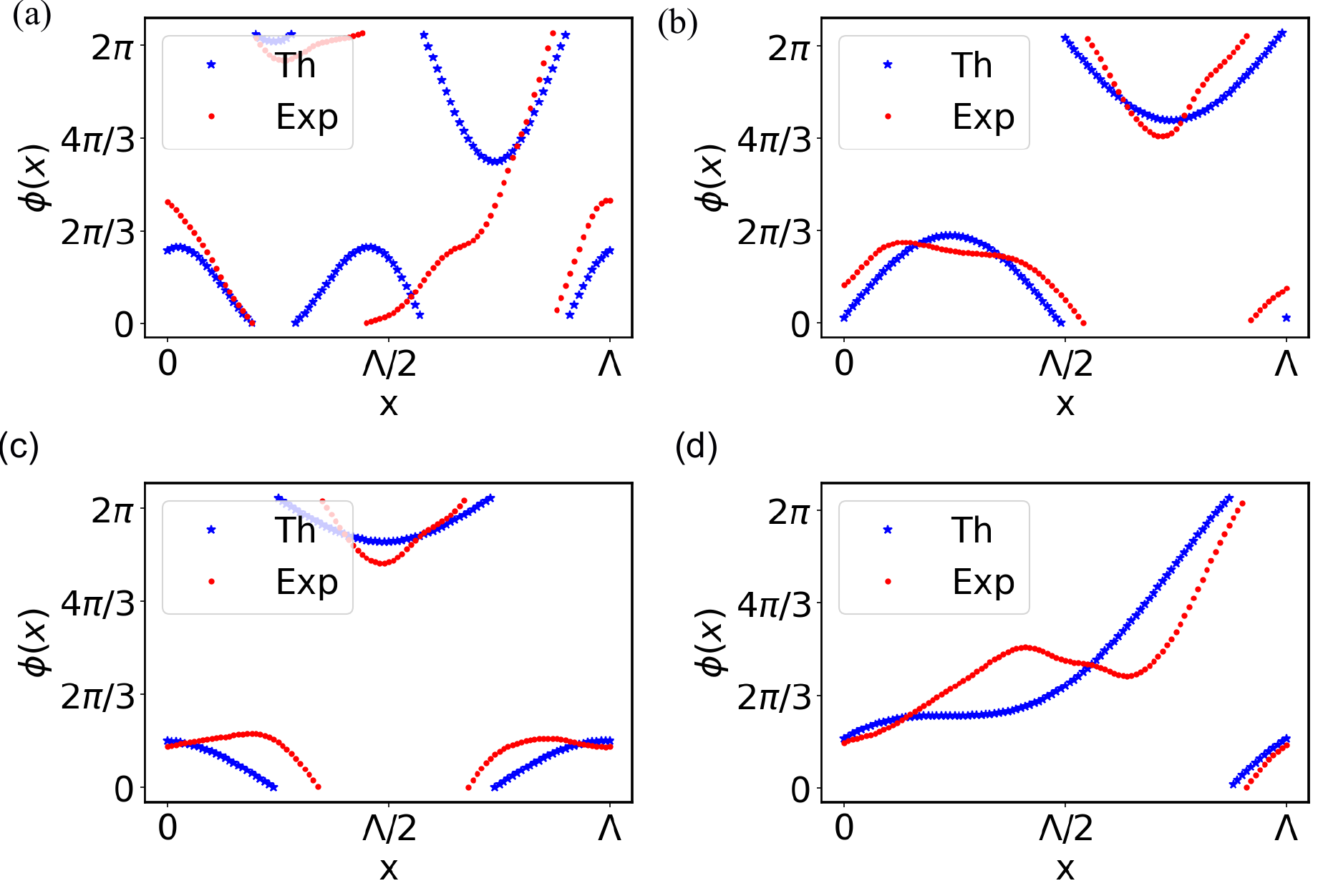}
  \caption{\textbf{Phase reconstruction of one-dimensional unitary operations.} Experimental phase reconstructions of the 1D lattice unitaries. (a) $\phi_1(x)=1.3\sin(\Delta k_{\perp}x)+1.5\cos(2\Delta k_{\perp}x)$, (b) $\phi_2(x)=1.9\sin(\Delta k_{\perp}x)$, (c) $\phi_3(x)=\cos(\Delta k_{\perp}x)$, and (d) ${\phi_4(x)=\cos(\Delta k_{\perp}x)+\Delta k_{\perp}x}$. 
  }
  \label{Fig:1DPhase}
\end{figure}

%
\begin{figure}[t]
  \centering
  \includegraphics[scale=0.38]{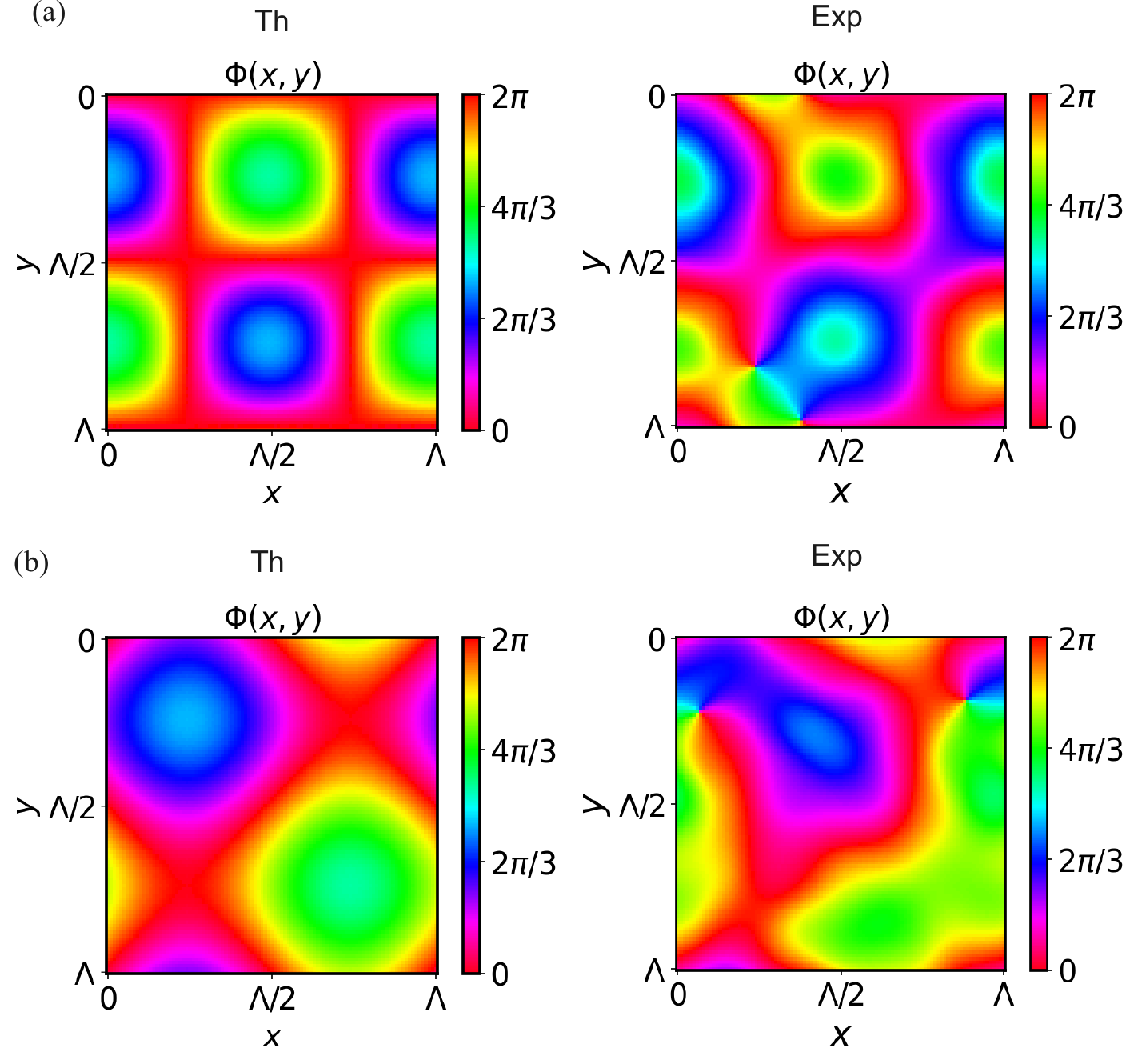}
  \caption{\textbf{Phase reconstruction of two-dimensional unitary operations.} Experimental phase reconstructions of the 2D lattice unitaries. (a)~$\phi_1(x,y)=2.8\sin{\left(\Delta k_\perp x\right)}\cos{\left(\Delta k_\perp y\right)}$ and (b)~$\phi_2(x,y)=1.4\sin{\left(\Delta k_\perp x\right)}+1.4\sin{\left(\Delta k_\perp y\right)}$.} 
  \label{Fig:2Dwalk}
\end{figure}

\subsection{Phase retrieval}
Since we use a phase device to implement our scheme, the unitaries we consider experimentally are of the form ${U(x,y) = e^{i\phi(x,y)}}$.
As an additional check of the quality of the transferred results, the phase transformations implementing the unitary action can be experimentally retrieved from two intensity distributions recorded in conjugate planes. In our case, we employ a hybrid Gerchberg-Saxton (GS) algorithm~\cite{GS}, which uses the near-field signal and far-field idler images (see Eq.~\eqref{eqn:nearfieldSLM2} and Eq.~\eqref{eqn:farfieldSLM1}, respectively), assuming perfect uniformity of the signal transverse profile. The GS algorithm is an iterative phase-retrieval method that reconstructs the phase of a complex wavefront from intensity measurements in two conjugate planes, typically the image and Fourier planes. Starting from a random input guess, it numerically propagates the beam back and forth between the two planes using the measured amplitudes as constraints, updating the phase at each iteration.
This provides a non-interferometric approach to phase reconstructions.
One of its strengths is that the error can only decrease (or stay the same) at each iteration \Cite{Fienup:82}. However, this strategy presents intrinsic limitations. In particular, the convergence speed is sensitive to the initial phase guess, and the convergence to a global minimum is not guaranteed. For these reasons, multiple runs of the algorithm are typically needed for optimal convergence, with the input phase guess randomized at each trial. This allows us to identify the optimal phase reconstruction (up to global shifts) as the one that minimizes the total distance (summed over all the pixels) between the reconstructed and measured near-field amplitude profiles. The number of repetitions is chosen to ensure the stability of the final reconstruction. Alternatively, the routine can be stopped when the algorithm has converged below a tolerance threshold. In our experiment, we set ${N_\text{R}=200}$ and ${N_\text{I}=200}$ for the 1D case, where ${N_\text{R}}$ is the number of independent runs and ${N_\text{I}}$ the number of iterations within each run, while for the 2D case we set ${N_\text{R}=100}$ and ${N_\text{I}=100}$. The total computation time is less than 5~s and about 80~s for the 1D and 2D implementations, respectively.

The reconstructed phase modulations are plotted in Fig.~\ref{Fig:1DPhase} and Fig.~\ref{Fig:2Dwalk} for the 1D and 2D implementations, respectively, where the comparison with the expected profile is also provided. For all reconstructions, a qualitatively good agreement is observed with the theoretical predictions, with some larger deviations in the 2D case that can be mainly ascribed to the low spatial resolution of the SLM and the camera, as well as aberrations in phase and amplitude of the pump beam, in addition to the intrinsic limitations of the GS algorithm.

\section*{Discussion and Conclusion}
We demonstrated the nonlocal transfer of unitary operations between correlated photons in high dimensions. 
In our scheme, the party with access to the computational resource can transfer the desired output to remote clients upon a suitable projective measurement. This operation is accomplished at the expense of high-dimensional spatial correlations. The technique has been experimentally validated in both 1D and 2D configurations in the case of phase transformations. 

Our setup efficiently processes a large number of co-propagating optical modes, which suggests potential use in future entanglement-based quantum key distribution~\cite{PhysRevLett.67.661} and quantum simulation protocols~\cite{Joshi2023}. 

The implementation of the proposed protocol with transverse momentum modes in real-world scenarios would certainly suffer from the typical challenges of free-space optical communications, namely, the presence of atmospheric turbulence. Adaptive optics systems offer an effective solution for mitigating the effects of intermediate-scale turbulence or for operation over moderate propagation distances \Cite{Hickson2014,scarfe:2025}. Conversely, one could design multi-plane light converters to compensate for aberrations induced by turbulence \Cite{10.1117/1.OE.61.11.116104}. Another approach could be to predict the turbulence strength in advance to find the best time to establish a secure connection \Cite{Jaouni:25}. Opposite to free-space solutions, multimode fibers could also be employed, but in that case, the distortion induced by the fiber itself must be taken into account and compensated for \Cite{Shen:05,Valencia2020}. Alternatively, since the biphoton state is also correlated in other degrees of freedom, such as frequency, one could adapt the protocol in such Hilbert spaces.

With high-dimensional correlations being the only physical requirement, the same scheme can also be applied to transfer operations in the orbital angular momentum space~\cite{mair2001entanglement}. By replacing the phase holograms with birefringent patterned optical elements, such as liquid-crystal~\cite{DiColandrea2023} or dielectric metasurfaces~\cite{Devlin:17}, our technique could be refined to transfer more complex operations coupling polarization and spatial degrees of freedom. Moreover, by adding a controlled amount of losses on a subset of modes, the extension to non-unitary transformations could also be explored~\cite{Wang:23}. Further exciting prospects involve the generalization of similar concepts to a larger number of input photons, typical in Boson sampling implementations~\cite{Zhong2020,Arrazola2021}. This will require further theoretical investigations.

\vspace{0.5cm}
\noindent\textbf{Acknowledgement.}  This work was supported by the Canada Research Chair (CRC) Program, NRC-uOttawa Joint Centre for Extreme Quantum Photonics (JCEP) via the Quantum Sensors Challenge Program at the National Research Council of Canada, and Quantum Enhanced Sensing and Imaging (QuEnSI) Alliance Consortia Quantum grant. FDC acknowledges support from the PNRR MUR project PE0000023-NQSTI.

\vspace{0.5cm}
\noindent \textbf{Disclosures.}
The authors declare no conflicts of interest.
\\
\newline \noindent \textbf{Data availability.} The source code and data underlying the results presented in this paper can be obtained from the corresponding author upon reasonable request.
\\
\newline \noindent \textbf{Author Contributions.}
DP and FDC conceived the idea. DP developed the theory for general unitary operators. FDC developed the theory adapted to translation-invariant operators and designed the holograms. DP and FDC performed the experiment and data analysis. DP and FDC prepared the first draft of the manuscript. ADE and EK supervised the project. All authors discussed the results and contributed to the final version of the manuscript.

\bibliography{bibliography}

\end{document}